\documentclass{emulateapj} 

\begin{document}

\slugcomment{submitted to ApJ}

\title{Gemini Spectroscopy of Ultra-Compact Dwarfs in the Fossil Group NGC 1132}


\author{Juan P. Madrid\altaffilmark{1} and Carlos J. Donzelli\altaffilmark{2,3}}

\altaffiltext{1}{Centre for Astrophysics and Supercomputing,
Swinburne University of Technology, Hawthorn, VIC 3122, Australia}

\altaffiltext{2}{Instituto de Investigaciones en Astronom\'ia Te\'orica y Experimental 
IATE, Observatorio Astron\'omico de  C\'ordoba,
Laprida 854, X5000BGR, C\'ordoba, Argentina}
\altaffiltext{3}{Consejo Nacional de Investigaciones Cient\'ificas y T\'ecnicas (CONICET),
Avenida Rivadavia 1917, C1033AAJ, Buenos Aires, Argentina}


\begin{abstract}

A spectroscopic follow up of  Ultra-Compact Dwarf (UCD) candidates in the fossil group
NGC 1132 is undertaken with  the Gemini Multi Object Spectrograph (GMOS). 
These new Gemini spectra prove the presence of six UCDs in the fossil 
group NGC 1132 at a distance of D$\sim$100 Mpc and a recessional velocity 
of $v_r = 6935 \pm 11$ km/s. The brightest and largest member of the UCD population
is an M32 analog with a size of 77.1 pc and a magnitude of M$_V$=-14.8 mag with the 
characteristics in between those of the brightest UCDs and compact elliptical galaxies. 
The ensemble of UCDs have an average radial velocity of $<~v_r~> = 6966 \pm 208$ km/s 
and a velocity dispersion of  $\sigma_v = 169 \pm 18$ km/s similar to the one of poor galaxy groups. 
This work shows that UCDs can be used as test particles to determine the dynamical 
properties of galaxy groups. The presence of UCDs in the fossil group environment 
is confirmed and thus the fact that UCDs can form across diverse evolutionary conditions.

\end{abstract}



\section{Introduction}

Ultra-Compact Dwarfs (UCDs) are compact stellar systems with characteristics 
between those of globular clusters and compact elliptical galaxies, particularly
with sizes between 10 and 100 pc. UCDs were discovered in the Fornax cluster
(Hilker et al.\ 1999; Drinkwater et al.\ 2000). Subsequent searches in 
other galaxy clusters revealed that, far from being an oddity to Fornax, UCDs 
were present in all major galaxy clusters in the nearby universe (Br\"uns \& Kroupa 2012
and their comprehensive set of references therein). 

In recent years several studies have searched for UCDs in environments other 
than galaxy clusters. New UCDs have been found in low density environments 
(Hau et al.\ 2009), galaxy groups (Romanowsky et al.\ 2009), and field galaxies 
(Norris \& Kannappan 2011). Within this framework, and through the analysis 
of Advanced Camera for Surveys imaging, Madrid (2011) found 11 UCDs 
and 39 extended star cluster candidates associated with the fossil 
group NGC 1132. These objects were identified through the analysis of their 
colors, luminosity, and structural parameters.

Fossil groups are galaxy systems where a luminous early type galaxy is brighter
than any other group member by more than two magnitudes. This dominant galaxy is believed
to be the end product of the merger of a galaxy group into a single entity (Ponman et al.\ 1994). 
Fossil groups have X-ray luminosities comparable to galaxy groups but a characteristic absence of 
$L^\star$ galaxies (Jones et al.\ 2003). Analysis of numerical simulations show
that fossil groups had an early assembly and a passive evolution thereafter (e.g.\ 
D\'iaz-Gim\'enez et al.\ 2011).  The surface brightness profile 
and the globular cluster system of the fossil group NGC 1132 were studied
by Alamo-Mart\'inez et al.\ (2012). These authors find that both surface brightness
and specific frequency of globular clusters ($S_N = 3.1 \pm 0.3$) in NGC 1132 
are similar to those of normal elliptical galaxies.   

Nine out of eleven UCD candidates found by Madrid (2011) share 
the same parameter space as the brightest globular clusters in the 
color magnitude diagram of the NGC 1132 globular cluster system. 
One UCD candidate, that will be designated as UCD1, is almost 
four magnitudes brighter than the brightest globular cluster 
associated with NGC 1132. UCD1 is 6.6 kpc from the center of NGC 1132,
has a half-light radius of 77.1 pc and a magnitude of $m_{F850LP} = 18.49$ mag.
A second UCD candidate, for which a spectrum was also obtained, had
particularly blue colors compared to the
 most luminous globular clusters.

The characteristics of UCD1 make it an interesting object since it 
is the link between the most massive UCDs found so far and the lowest 
mass M32-like galaxies. UCD1 is the brightest object in a recent compilation 
of 813 UCDs and extended objects (Br\"uns \& Kroupa 2012). Indeed, high surface 
brightness compact elliptical galaxies, or “M32-like” stellar systems are rare. 
Chilingarian \& Mamon (2008) list only six confirmed stellar systems similar to M32. 

As pointed out recently by Br\"uns \& Kroupa (2012) only one fifth of all 
published Ultra-Compact Dwarfs have been so far spectroscopically confirmed. 
In this work, we present the results of an observing campaign with 
Gemini North to obtain spectroscopic follow-up  of  UCD candidates 
in the fossil group NGC 1132 presented by Madrid (2011).


\section{Observations} 

Spectra for seven UCD candidates, and the brightest globular cluster
in the fossil group NGC 1132 were obtained with the Gemini North telescope 
using the Gemini Multi Object Spectrograph (GMOS). These observations
were acquired under program GN-2012B-Q-10.
UCD candidates were determined through the analysis of HST imaging. 
A multislit mask was created using a pre-image provided by Gemini. 
Given that the Advanced Camera for Surveys on board HST has a field of view of
$202\times202$ arcseconds all targets  are located within 
this small area centered on the host galaxy. The central concentration 
limited the total number of targets that fit on the GMOS slit mask.

The spectroscopic data were acquired in queue mode on 20 September 2012 using a multislit 
mask. Individual slits had a width of one arcsecond. The grating in use was the B600$+_{-}$G5323 
that has a ruling density of 600 lines/mm. Three exposures of 1800 s each were obtained with 
the central wavelengths of 497, and 502 nm. Two additional exposures of 1800 s each were 
obtained with a central wavelength of 507 nm. Science targets have thus a total exposure 
time of $\sim$4 hours. Flatfields, spectra of the standard star $BD +28~4211$, and the 
copper-argon $CuAr$ lamp were also acquired to perform flux calibration. A binning 
of $2\times 2$ was used, yielding a scale of 0.1456 arcseconds per pixel and a theoretical 
dispersion of $\sim 0.9$ \AA~ per pixel. Our targets transit $\sim30\degr$ 
from the zenith and if the position angle is set to $90\degr$ the displacement due to
atmospheric refraction is small in our configuration. Slit losses due to atmospheric 
refraction can be neglected. 


\section{Data Reduction and Analysis}

All science and calibration files were retrieved from the Gemini Science Archive 
hosted by the Canadian Astronomy Data Center. The data reduction described below
was carried out with the Gemini IRAF package. Flatfields were derived with the
task {\sc gsflat} and the flatfield exposures. Spectra were reduced using 
{\sc gsreduce}, this does a standard data reduction, that is, performs bias, 
overscan, and cosmic ray removal as well as applying the flatfield 
derived with {\sc gsflat}. GMOS-North detectors are read with six 
amplifiers and generates files with six extensions. The task {\sc gmosaic} 
was used to generate data files with a single extension. The sky level was 
removed interactively using the task {\sc gskysub} and the spectra were 
extracted using {\sc gsextract}.

Flux calibration was performed using the spectra of the standard star $BD +28 4211$, 
acquired with an identical instrument configuration. Spectra of $CuAr$ lamps were 
obtained immediately after the science targets were observed and were used 
to achieve wavelength calibration using the task  {\sc gswavelength}. 
We use {\sc gstransform} to rectify, interpolate, and calibrate the spectra using the 
wavelength solution found by {\sc gswavelength}. The sensitivity function of the 
instrument was derived using {\sc gsstandard}  and the reference file for 
$BD +28 4211$ provided by Gemini observatory. Science spectra were flux 
calibrated with {\sc gscalibrate} which uses the sensitivity function 
derived by {\sc gsstandard}.


\begin{figure*}
\epsscale{1.1}
\plotone{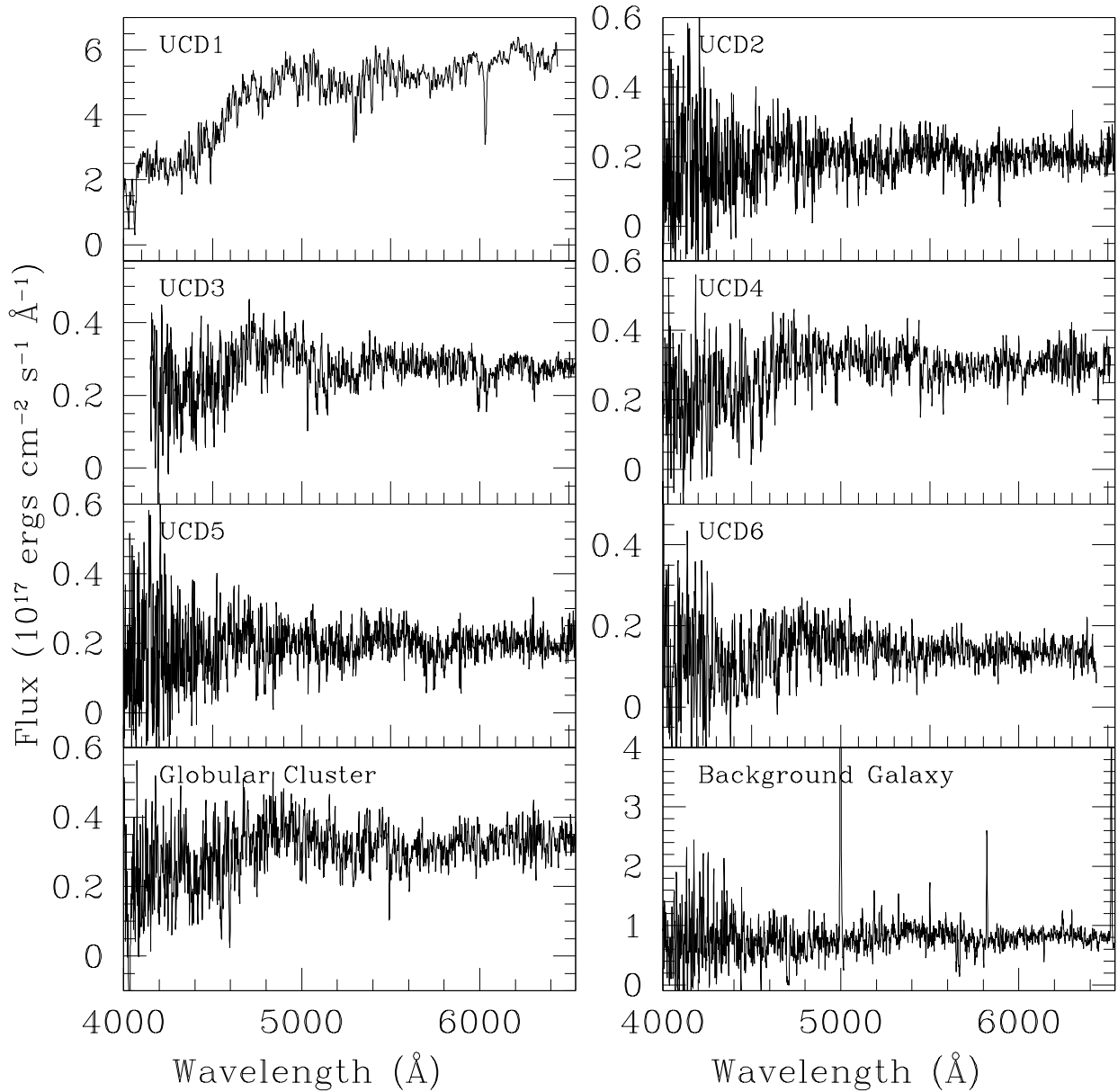}
\caption{Calibrated Gemini/GMOS spectra of six confirmed UCDs, one globular cluster
and a background galaxy in the NGC 1132 field. The most prominent lines of these
spectra, particularly UCD1, are the magnesium and sodium doublets with 
restframe wavelengths of 5169, 5175 \AA~ and  5890, 5896 \AA~  respectively. 
Note that these spectra are not corrected for radial velocities.
 \label{fig1}}
\end{figure*}



\begin{figure}
\epsscale{1.25}
\plotone{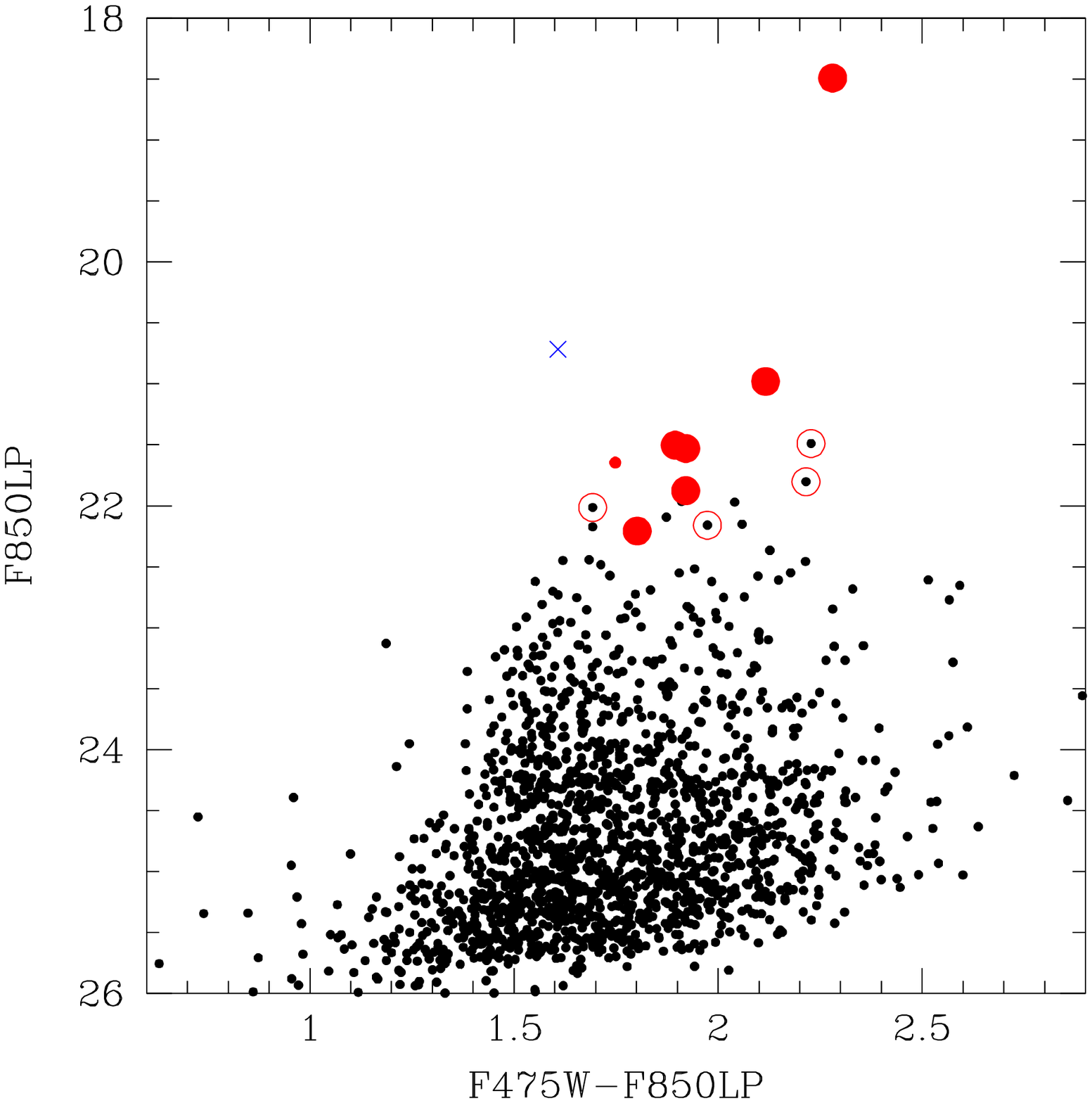}
\caption{Color-Magnitude Diagram of the globular cluster system of NGC 1132
(black dots). UCD candidates are represented as open red circles. Objects with membership 
to NGC 1132 confirmed with the Gemini data presented in this paper 
are displayed as solid red circles. The small red point is GC1. The  blue cross represents a background galaxy that 
masqueraded as an UCD candidate with blue colors in the HST images. This CMD 
was originally derived and presented by Madrid (2011). A detailed analysis of 
the globular cluster system of NGC 1132 is given by Alamo-Mart\'inez et al.\ (2012). 
 \label{fig2}}
\end{figure}



\section{Results}

Figure 1 presents the calibrated spectra of six UCDs, the brightest 
globular cluster of the NGC 1132 globular cluster system, and a 
background galaxy thought to be a UCD with particularly blue colors.
Various absorption lines are identifiable in the spectra of UCDs, 
particularly in the spectrum of UCD1 which has a higher signal to noise 
ratio. The two most prominent absorption features in the spectrum of UCD1
are the magnesium and sodium doublets with restframe wavelengths 
of 5169, 5175 \AA~ and  5890, 5896 \AA~ respectively. These two doublets 
are also evident in the UCD spectra published by Francis et al.\ (2012).
Other lines present in the spectrum of UCD1 are Ca and Fe (5269 \AA),
Fe (5331 \AA; 4384), HeII (5411 \AA), H$\beta$ (4861 \AA), and 
CaI (4227 \AA).



\subsection{Redshift Determination}

The redshift of the targets is derived using the 
IRAF task {\sc fxcor} that computes radial velocities by deriving the Fourier 
cross correlation between two spectra. As a reference spectrum we use data
of the Galactic globular cluster BH 176 taken in the same GMOS configuration
during a previous Gemini run (Davoust et al.\ 2011). For UCD1 the 
task {\sc fxcor} returns a radial velocity of 7158 $\pm$32 km/s.
The host galaxy NGC 1132 has a published radial velocity of 6935$\pm$11 km/s   	
(Collobert et al.\ 2006). With these results we confirm that UCD1 
is located within the fossil group NGC 1132 at a redshift of $z\sim$0.023.
This redshift measurement validates the photometric and structural parameters
derived by Madrid (2011) for UCD1. 


\subsection{Age and Metallicity of UCD1, an M32-like Object}

A caveat for this section is relevant: an accurate age and metallicity 
determination of the old stellar systems studied here is difficult. 
In a similar spectroscopic study of UCDs, Francis et al.\ (2012) derived 
the ages of 21 Ultra-Compact Dwarfs using both Lick indices and spectral fitting
of Simple Stellar Populations and the results of these two techniques do not 
correlate with one another. The metallicities of these 21 UCDs, also 
derived by  Francis et al.\ (2012), have better uncertainties and show 
a correlation between the two methods with an offset of 0.2 dex.

The stellar populations synthesis code {\sc starlight} (Cid Fernandes et al.\ 2005)
was used with the aim of deriving the metallicity of UCD1. {\sc starlight} compared 
the Gemini spectrum of UCD1 with a database of 150 spectral templates and found that the best
fit is provided by a combination of two stellar populations: one representing 30\% 
of the flux and having solar metallicity ($Z$=0.02) and  a second population 
accounting for 70\% of the flux and having supersolar metallicity ($Z=0.05$).
This spectral fitting also yields an age of 13 Gyr for UCD1.


For spectra with sufficient signal-to-noise Lick indices (Worthey et al.\ 1994)
can be derived. The code GONZO (Puzia et al.\ 2002) was used to derive Lick
indices for UCD1 which is the only UCD with enough signal to 
generate significant results. The Gemini spectrum is degraded to the
Lick resolution by GONZO. We also apply the zero points of the calibration given 
by Loubser et al. (2009). The values for the lick indices and their associated Poisson errors 
are given in Table 1, this allows comparison with existent and future studies. By comparing 
the indices $<Fe>$ and Mgb to the single stellar population models of Thomas et al.\ (2003) 
we can derive an $\alpha$-element abundance for UCD1 of $[\alpha/ \rm Fe]$=+0.3. Indices H$\beta$=1.96 \AA~ and Fe 5270 = 3.57 \AA~ of 
UCD1 are very similar to the values derived by Chiboucas et al.\ (2011) for 
UCD 121666 in the Coma cluster. UCD 121666 has an H$\beta$=1.84 \AA~ and Fe 5270 = 3.48 \AA.

The Lick Indices we obtained were given as input to the publicly available code
EZ Ages (Graves \& Schiavon 2008). This code determines ages and abundances of 
unresolved stellar populations using their Lick indices. We chose an alpha-enhanced
isochrone fitting  and found an age of 7.5 Gyr and an iron abundance of $[Fe/H]$= -0.17.
The chemical abundances of  Grevesse \& Sauval (1998) yield a metallicity of $Z=0.015$.
The metallicity derived for UCD1 is high among UCDs but not unprecedented (Chiboucas et al.\ 2011; 
Francis et al.\ 2012).


\begin{deluxetable*}{lccccccc}
\tablecaption{Lick Indices for UCD1\label{tbl-1}} 
\tablehead{
\colhead{$H\beta$} & \colhead{$H\delta$A} & \colhead{$H\gamma A$} &  \colhead{Mgb} & \colhead{Fe 5270} &  \colhead{Fe 5335}  & \colhead{$<Fe>$} &\colhead{[MgFe]} 
}
 
\startdata

1.96 $\pm 0.02$ & -1.92 $\pm 0.15$ & -7.69 $\pm 0.15$ & 5.03 $\pm 0.02$ & 3.57 $\pm 0.02$ & 1.86 $\pm 0.01$ & 2.7 $\pm 0.02$ & 3.1 $\pm 0.03$\\

 \enddata

\end{deluxetable*}


\subsection{Internal Velocity Dispersion}

The main objective of this work is to obtain redshifts for the UCD candidates.
Determining their internal velocity dispersion, roughly of a few tens of 
kilometers per seconds is beyond the resolution of these Gemini/GMOS observations. 
The empirical spectral resolution of our observations has a FWHM$\sim200$ km/s at 
5000\AA, that is about 10 times the value of published internal velocity dispersion of UCDs (e.g.\  $\sigma_v=20.0$ km/s Hasegan et al.\ 2005). 


\subsection{UCD Population in the Fossil Group NGC 1132}

The redshifts derived with these new Gemini spectra confirm six UCD
candidates as true members of the fossil group NGC 1132. The redshifts, 
photometric, and structural parameters of these  UCDs are listed in Table 2.
UCD2 through UCD6 are the extension to higher luminosities and redder colors of 
the brightest globular clusters of NGC 1132. The sizes of UCD2 through UCD6 
range  between 9.9 and 19.8 pc and all globular clusters have a size smaller 
than $\sim$8 pc. The Color-Magnitude Diagram (CMD) of the NGC 1132 globular 
cluster system is plotted in Figure 2. This CMD also contains the colors and 
magnitudes of UCD candidates and objects with spectroscopic data.

These six UCDs have an average radial velocity of $<~v_r~> = 6966 \pm 208$ 
km/s. The velocity dispersion of this family of UCDs is  $\sigma = 169 \pm 18$
km/s which is in the range of poor galaxy groups, as discussed below. 
The velocity dispersion was derived, through bootstrapping, using the 
prescriptions of Strader et al.\ (2011, their Equation 4). 
UCDs can be used in the same manner as globular clusters and planetary nebulae 
have been used to trace the dynamics of nearby galaxies (e.g. Coccato et al.\ 2009).


\begin{center}
\begin{deluxetable*}{lcccccccc}
\tablecaption{Radial Velocities, Photometric and Structural Parameters\label{tbl-1}} 
\tablehead{
\colhead{ID} & \colhead{RA} & \colhead{Dec} & \colhead{Radial Velocity} &  \colhead{$m_{F850LP}$} & \colhead{$M_{F850LP}$}   & \colhead{Color} &\colhead{$r_h$ (pc)} &\colhead{$R_{GC}$ (kpc)}
}
 
\startdata

UCD1  & 2h 52m 51s & $-1\degr 16\arcmin 19\arcsec$ & 7158 $\pm$ 32  & 18.5 & -16.4 & 2.28 & 77.1 & 6.6\\

UCD2  & 2h 52m 54s & $-1\degr 17\arcmin 39\arcsec$ & 6834 $\pm$ 84 & 21.0  & -13.9 & 2.12 & 13 .3  & 36.0\\

UCD3  & 2h 52m 59s & $-1\degr 14\arcmin 54\arcsec$ & 7082 $\pm$ 69 & 21.5  & -13.4 & 1.89 & 19.8  & 65.5\\

UCD4  & 2h 52m 52s & $-1\degr 15\arcmin 43\arcsec$ & 6627 $\pm$ 60 & 21.5  & -13.3 & 1.92 & 16.7  & 21.7\\

UCD5  & 2h 52m 55s & $-1\degr 17\arcmin 19\arcsec$ & 6949 $\pm$ 120 & 21.9 & -13.0 & 1.92 & 13.0 & 32.3\\

UCD6  & 2h 52m 51s & $-1\degr 16\arcmin 06\arcsec$ & 7147 $\pm$ 140 & 22.2 & -12.7 & 1.80 & 9.9  & 12.0\\

GC1   & 2h 52m 53s & $-1\degr 16\arcmin 34\arcsec$ & 6948 $\pm$ 99  & 21.7 & -13.2 & 1.75 & --   & 5.6\\

 \enddata

 \tablecomments{Column 1: Identifier; Column 2: Right Ascension; Column 3: Declination; Column 4: radial velocity;
 Column 5: apparent magnitude in the HST filter F850LP (similar to Sloan $z$); Column 6: absolute magnitude; Column 7: 
 color (F474W-F850LP); Column 8: effective radius in parsecs. The size of GC1 is below the resolution limit of HST, 
 in this case $\sim$8 pc in radius; Column 9: projected distance to the center of NGC 1132, in kpc.}

\end{deluxetable*}
\end{center}


\subsection{A Blue UCD Candidate}

In Madrid (2011) a particularly ``blue" Ultra-Compact Dwarf candidate was reported.
This UCD candidate satisfied all selection criteria based on size, ellipticity,
magnitude, and color. This candidate was the bluest UCD candidate and seemingly
did not follow the Mass-Metallicity Relation (MMR) relation of massive globular
clusters (M$> 10^{6} M_{\odot}$) (Bailin \& Harris 2009). A Gemini-GMOS spectrum of this
object reveals that it is actually a background star-forming galaxy with 
strong emission lines. This object is represented by a cross on the CMD 
in Figure 2. Madrid (2011) showed that UCDs have, on average, redder colors than
extended globular clusters.

\subsection{Spatial Distribution of UCDs}

The spatial distribution of globular clusters, UCD candidates, and UCDs
with spectroscopic confirmation is presented in Figure 3. Both globular
clusters and UCDs appear to follow the same spatial distribution. UCDs and
globular clusters aggregate toward the center of the host galaxy as is
expected for these satellite systems. There is no other particular clustering
or alignment of these objects in the ACS field. The projected galactocentric 
distances of UCDs are given in Table 2. At 6.6 kpc UCD1 is the closest 
UCD to the center of NGC 1132.


\begin{figure}
\epsscale{1.25}
\plotone{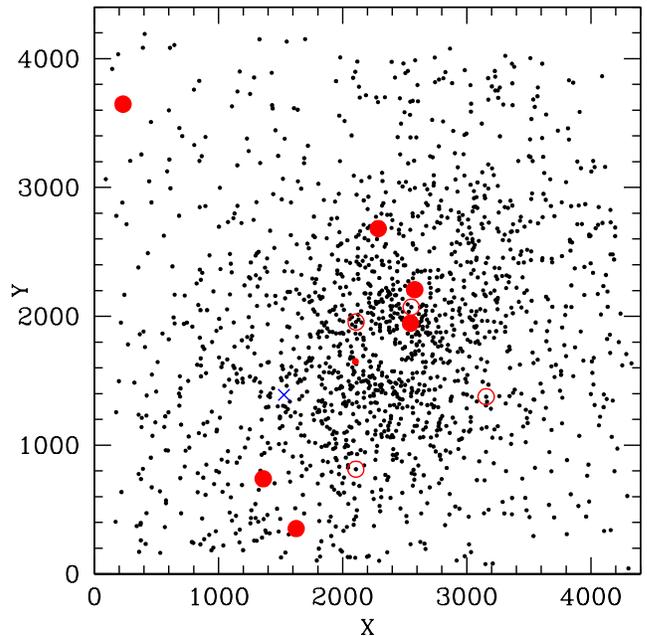}
\caption{Spatial distribution of globular clusters (black dots), UCD candidates
(open red circles), and UCDs with spectroscopic confirmation (solid red circles).
The small red dot is GC1. The background galaxy with photometric properties indistinguishable from UCD
candidates is represented by a blue cross. This figure represents the field of view
of the Advanced Camera for Surveys that, at the distance of NGC 1132, corresponds to  
$\sim 100 \times 100$ kpc.
 \label{fig2}}
\end{figure}


\section{Discussion}

\subsection{Properties of the Fossil Group NGC 1132}

Mulchaey \& Zabludoff (1999) call for the use of dwarf galaxies 
as test particles for the study of the dynamics and dark matter halo
of NGC 1132. UCDs can fulfill this role. In galaxy groups 
$M^{\star}$ galaxies are expected to merge in a fraction of a 
Hubble time while on the other hand dwarf galaxies have dynamical 
friction timescales greater than a Hubble time  (Mulchaey \& Zabludoff 1999).
Albeit based on a small sample, the velocity dispersion of the 
UCD population can be used in existing scaling relations for the X-ray 
luminosity and group richness. 

Several studies have derived a relation between X-ray luminosity ($L_X$)
and velocity dispersion $\sigma_v$ of galaxy clusters (Ortiz-Gil et al.\ 2004 and 
references therein), and galaxy groups (Xue \& Wu 2000). Whether fossil galaxy groups
follow the same scaling relations of galaxy clusters or have a shallower 
$L_X \propto\sigma_v $ relation is still a matter of debate (Khosroshahi et al.\ 2007).

A relation between X-ray luminosity ($L_X$) and velocity dispersion $\sigma_v$
for galaxy groups is $L_X~=~10^{-2.95\pm0.30}\sigma_v^{1.00\pm0.12}$ in units of 
10$^{42}$ erg/s (Xue \& Wu 2000). If we insert the value for  the velocity 
dispersion of UCDs $\sigma_v = 169$ km/s in the above formula the X-ray luminosity 
is  $L_X = 0.2 \times 10^{42}$. This value is more than a factor of ten lower than the value
reported by Mulchaey \& Zabludoff (1999) of $\sim 2.5\times 10^{42} h^{-2}_{100}$ erg/s.
A better match to the observation is given by the relation between $L_X$ and $\sigma$ for 
galaxy clusters derived by Ortiz-Gil et al.\ (2004). These authors give the following
relation for what they call a volume limited sample: $L_X =10^{35.16\pm 0.09} \sigma^{3.2\pm0.3}$.
This formula yields  $L_X = 1.9 \times 10^{42}$ erg/s, which is a better approximation
to the observed value. In a recent work, Connelly et al.\ (2012) make  a detailed 
analysis of a sample of galaxy groups and give $L_X-\sigma$ relations that depend on
different factors, for instance,  number of group members and radial cuts applied. 
One of the Connelly et al.\ (2012) relations yields an X-ray luminosity 
of $L_X \sim 3 \times 10^{43}$ erg/s. There is a discrepancy between the different 
$L_X-\sigma$ relations published in previous studies. 

Pisani et al.\ (2003) derived a correlation between group richness, or the number 
of group members ($N$), and velocity dispersion ($\sigma_v$): $\log N = 127 \log \sigma_v -1.47$.
If the velocity dispersion of UCDs are indicative of the primordial velocity 
dispersion of NGC 1132,  according to the  Pisani et al.\ (2003) formula
the  UCD population has the velocity dispersion corresponding to  
a poor group with  $N=22$ members. 

Mendes de Oliveira \& Carrasco (2007) report the velocity dispersion of, among other 
objects, two Hickson Compact Groups (HCG) at low redshift: HCG 31 has a velocity
dispersion of $\sigma_v=60$ kms/s and HCG 79 has a velocity dispersion of $\sigma_v$=138 km/s. 
At $\sigma_v$=169 km/s the velocity dispersion of NGC 1132 is higher than these two 
low redshift compact groups but lower than the average $\sigma_v$=300 km/s found for 
a collection of 20 groups by Mahdavi et al.\ (1999).

\subsection{UCDs in Different Environments}

Analysis of high resolution HST data is a very efficient method to discover 
new Ultra-Compact Dwarfs in very diverse environments. The combined study of 
luminosity, colors, and structural parameters, possible with high 
resolution imaging data, yields UCD candidates with very high spectroscopic confirmation 
rates as shown in this work and in the study of the Coma cluster (Madrid et al.\ 2010, 
Chiboucas et al.\ 2011) among others. Searching for UCDs with seeing-limited data 
is a more arduous and unfruitful task (e.g.\ Evstigneeva et al.\ 2007).
One drawback of HST detectors is their small field of view that only covers
the innermost regions of host galaxies where the strongest tidal effects
take place. This work along with the publications cited in the introduction 
show that UCDs can be formed in environments with different evolutionary 
histories.


A historical scarcity of compact stellar systems with characteristic scale sizes 
between 30 and 100 parsecs created an unmistakable gap in size-magnitude 
relations between dwarf ellipticals, compact ellipticals and globular clusters. 
Gilmore et al.\ (2007) have interpreted the gap in the parameter space defined by 
compact stellar systems as a sign of two distinct families of objects, reflecting 
the intrinsic properties of dark matter. Globular clusters would belong to a family 
of dark matter-free stellar systems while dwarf spheroidals and compact ellipticals form the 
branch where dark matter is present or even dominant. Gilmore et al. (2007) postulate 
that dark matter halos have cored mass distributions with characteristic scale sizes 
of more than 100 pc. UCD1, however, with an effective radius of 77.1 pc is precisely in 
this gap of compact stellar systems. The brightest UCDs are the ideal candidates
to bridge the gap between compact ellipticals and globular clusters. 
Part of the gap is due to artificial selection 
effects introduced to eliminate contaminants in photometric studies. 
For instance, in the ACS Virgo Cluster Survey an upper limit of 10 
parsecs was imposed on globular cluster candidates.
Recent reanalysis of data lifting the 10 parcsecs upper limit on size 
for compact stellar systems have uncovered new systems previously 
ignored (e.g. Brodie et al.\ 2011 for the case of M87).

UCDs are believed to be the bright and massive tail of the globular
cluster luminosity function (Drinkwater et al.\ 2000) or the nuclei 
of stripped dwarf galaxies (Bekki et al.\  2003). A combination of both
formation mechanisms has also been proposed (Da Rocha et al.\ 2011).
The large magnitude gap between UCD1 and the brightest globular clusters 
of NGC 1132 suggests that UCD1 is the leftover core of a spiral galaxy 
(Bekki et al.\ 2001). The other UCDs can be stripped dE nuclei or massive globular clusters. The analysis of a large sample of spectroscopic properties of UCDs would prove a link between their stellar populations and those of dwarf galaxies
and/or globular clusters. 



\acknowledgments

Based on observations obtained at the Gemini Observatory, which is operated 
by the Association of Universities for Research in Astronomy, Inc., under a 
cooperative agreement with the NSF on behalf of the Gemini partnership: 
the National Science Foundation (United States), the Science and Technology Facilities
Council (United Kingdom), the National Research Council (Canada),
CONICYT (Chile), the Australian Research Council (Australia),
Minist\'{e}rio da Ci\^{e}ncia, Tecnologia e Inova\c{c}\~{a}o (Brazil)
and Ministerio de Ciencia, Tecnolog\'{i}a e Innovaci\'{o}n Productiva
(Argentina). The Gemini data for this paper were obtained under program GN-2012B-Q-10. 

We thank the referee for a prompt report with several comments that helped
to improve this paper. J.\ Madrid is grateful for a travel grant from the 
Australian Nuclear Science and Technology Organization (ANSTO). The access
to major research facilities is supported by the Commonwealth of Australia
under the International Science Linkages Program. Many thanks to V.\ Pota, 
J.\ Hurley, J.\ Cooke, E. Caris, A. Hou, P. Jensen, and L.\ Vega for enlightening 
discussions. This research has made use of the NASA Astrophysics Data System bibliographic 
services (ADS), the NASA/IPAC Extragalactic Database (NED), 
the SIMBAD database, and Google.


{\it Facilities:} \facility {Gemini North (GMOS); HST(ACS)}





\newpage

\begin{thebibliography}{}

\bibitem[Alamo-Martinez(2012)]{bai09} Alamo-Mart\'inez, K. A., West, M. J., Blakeslee, J. P. et al. 2012, A\&A, 546, 15

\bibitem[Bailin(2009)]{bai09} Bailin, J. \& Harris, W. E. 2009, ApJ, 695, 1082

\bibitem[Bekki et al.(2001)]{bek01} Bekki, K., Couch, W. J., Drinkwater, M. J., \&  Gregg,
M. D. 2001, ApJL, 557, L39

\bibitem[Bekki et al.(2003)]{bek03} Bekki, K., Couch, W. J., Drinkwater, M. J., \&  Shioya, Y. 2003, MNRAS, 344, 399

\bibitem[Brodie et al.(2011)]{brodie} Brodie, J. P., Romanowsky, A. J., Strader, J., Forbes, D. A. 2011, AJ, 142, 199

\bibitem[Br\"uns \& Kroupa(2012)]{bruns} Br\"uns, R. C. \& Kroupa, P. 2012, A\&A, 547, 65

\bibitem[Collobert et al.(2006)]{col06} Collobert, M., Sarzi, M., Davies, R. L.,
        Kuntschner, H. \& Colless, M. 2006, MNRAS, 370, 1213

\bibitem[Chiboucas et al.(2011)]{chi11} Chiboucas, K., Tully, R. B., Marzke, R. O. et al. 2011, ApJ, 737, 86

\bibitem[Chilingarian(2008)]{chl10} Chilingarian I. V. \& Mamon, G. A. 2008, MNRAS, 385, L83

\bibitem[Cid Fernandes(2005)]{cid2005} Cid Fernandes, R., Mateus, A., Sodr\'e, L., Stasinzka, G., \& Gomes, J. M., 2005, MNRAS, 358, 363 

\bibitem[Cocatto et al.(2009)]{cocatto09} Coccato, L., O. Gerhard, M., Arnaboldi, P. Das, et al. 2009, MNRAS, 394, 1249

\bibitem[Connelly et al.(2012)]{connelly12} Connelly, J. L., Wilman, D. J., Finoguenov, A. et al. 2012, ApJ, 756, 139

\bibitem[Da Rocha et al.(2011)]{darocha11}Da Rocha, C., Mieske, S., Georgiev, I. Y. et al. 2011, A\&A, 525, 86

\bibitem[Davoust et al.(2011)]{davoust11} Davoust, E., Sharina, M. E., \& Donzelli, C. J. 2011, A\&A, 528, 70

\bibitem[D\'iaz-Gim\'enez et al.(2011)]{diaz11} D\'iaz-Gim\'enez, E., Zandivarez, A., Proctor, R., Mendes de Oliveira, C., \& Abramo, L. R.
         2011, A\&A, 527, A129
        
\bibitem[Drinkwater(2000)]{dri00} Drinkwater, M. J. et al.\ 2000, PASA, 17, 227        

\bibitem[Evstigneeva et al.(2007)]{evs07} Evstigneeva, E. A., Drinkwater, M. J., 
Jurek, R., Firth, P., Jones, J. B., Gregg, M. D., \& S. Phillipps, S. 2007, MNRAS, 378, 1036

\bibitem[Francis et al.(2012)]{francis12} Francis, K. J., Drinkwater, M. J., Chilingarian, I. V., Bolt, A. M. \& Firth, P 2012, MNRAS, 425, 325

\bibitem[Gilmore et al.(2007)]{gilmore07} Gilmore, G., et al. 2007, ApJ, 663, 948 

\bibitem[Graves et al.(2008)]{graves08} Graves, G. \& Schiavon, R. P. 2008, ApJS, 177, 446

\bibitem[Grevesse \& Sauval(1998)]{grevesse98} Grevesse, N. \& Sauval, A. J. 1998, Space Science Reviews, 85, 161

 
\bibitem[Hasegan et al.(2005)]{Has05} Hasegan, M., et al. 2005, ApJ, 627, 203

\bibitem[Hau(2009)]{Hau09} Hau, G. K., et al.\ 2009, MNRAS, 394, L97 

\bibitem[Hilker(1999)]{hil99}Hilker, M., et al.\ 1999, A\&AS, 134, 75

\bibitem[Jones et al.(2003)]{jones03} Jones, L. R., Ponman, T. J., Horton, A., Babul, A., Ebeling, H. \& Burke, D. J. 2003, MNRAS, 343, 627

\bibitem[Khosroshahi et al.(2007)]{kho07} Khosroshahi, H. G., Ponman, T. J. \& Jones, L. R. 2007, MNRAS, 377, 595

\bibitem[Loubser et al. (2009)]{mad10} Loubser, S. I., S\'anchez-Bl\'asquez, P., Sansom, A. E., \& Soechting, I. K. 2009, MNRAS, 398, 133

\bibitem[Madrid(2011)]{mad11} Madrid, J. P. 2011, ApJL, 737, L13

\bibitem[Madrid et al. (2010)]{mad10} Madrid, J. P., et al. 2010, ApJ, 722, 1707 

\bibitem[Mahdavi et al. (1999)]{mahdavi99} Mahdavi, A., Geller, M. J., Bohringer, H., Kurtz, M. J. \& Ramella, M. 1999, ApJ, 518, 69

\bibitem[Mendes de Oliveira \& Carrasco(2007)]{mendes07} Mendes de Oliveira, C. L. \& Carrasco, E. R. 2007, ApJ, 670, L93

\bibitem[Mulchaey \& Zabludoff(1999)]{mul99} Mulchaey, J. S. \& Zabludoff, A. I. 1999, ApJ, 514, 133

\bibitem[Norris \& Kannappan(2011)]{nor11} Norris, M. A. \& Kannappan, S. J. 2011, MNRAS, 414, 739

\bibitem[Ortiz-Gil et al.(2004)]{ortiz04} Ortiz-Gil, A., Guzzo, L., Schuecker, P., B\"oringer, H. \& Collins, C. A. 2004, MNRAS, 348, 325

\bibitem[Pisani et al.(2003)]{pis03} Pisani, A., Ramella, M. \& Geller, M. 2003, AJ, 126, 1677

\bibitem[Ponman et al.(1994)]{pon94} Ponman, T. J., Allan, D. J., Jones, L. R., Merrifield, M., McHardy, I. M., Lehto, H. J. \& Luppino, G. A. 1994, Nature, 369, 462

\bibitem[Puzia et al.(2002)]{puz02} Puzia, T. H., Zepf, S. E., Kissler-Patig, M., Hilker, M., Minniti, D. \& Goudfrooij, P. 2002, A\&A, 391, 453

\bibitem[Romanowsky et al.(2009)]{roman09} Romanowsky, A. J., Strader, J., Spitler, L. R., Johnson, R.,  Brodie, J. P., Forbes, D. A., Ponman, T. 2012, AJ, 137, 4956  

\bibitem[Strader et al.(2006)]{stra06} Strader, J., Romanowsky, A. J., Brodie, J. P., Spitler, L. R. et al.\ 2011, ApJS, 197, 33  
     
\bibitem[Thomas et al.(2003)]{thomas03} Thomas, D., Maraston, C., Bender, R. 2003, MNRAS, 339, 897   
        
\bibitem[Worthey et al.(1994)]{worthey94} Worthey, G., Faber, S. M., Gonzalez, J. J. \& Burstein, D. 1994, ApJS, 94, 687
   
\bibitem[Xue \& Wu(2000)]{xue00} Xue, Y-J. \& Wu, X-P 2000, ApJ, 538, 65
    

\end{thebibliography}
\end{document}